\documentclass{amsart}

\usepackage[latin1]{inputenc}

\usepackage[dvips]{graphics}
\graphicspath{{Fig/}}

\newtheorem{definition}{Definition}

\def\etal{{\em et al.}}

\title[Adaptive algorithms for IP Traffic]{Adaptive algorithms for the identification of large flows in IP traffic}

\author{Youssef Azzana}\thanks{Work done during PhD preparation at INRIA Paris -- Rocquencourt}
\address[Y. Azzana]{Eurafric Information,
Casablanca, Marrocco}

\author[Chabchoub]{Yousra Chabchoub}
\address[Y. Chabchoub, C. Fricker, Ph. Robert]{INRIA Rocquencourt,  RAP
    project, Domaine de Voluceau, 78153 Le Chesnay, France. }
\email{Yousra.Chabchoub@inria.fr}

\author[Fricker]{Christine Fricker}
\email{Christine.Fricker@inria.fr}

\author[Guillemin]{Fabrice Guillemin}
\address{France Telecom, Division R\&D, 2 Avenue Pierre Marzin, 22300 Lannion, France}
\email{Fabrice.Guillemin@orange-ft.com}

\author[Robert]{Philippe  Robert}
\email{Philippe.Robert@inria.fr}

\begin{document}
\maketitle
\begin{abstract}
We propose in this paper an on-line algorithm based on Bloom filters for identifying large
flows in IP traffic  (a.k.a. elephants). Because of the large number  of small flows, hash
tables of these algorithms have to  be regularly refreshed.  Recognizing that the periodic
erasure scheme usually used in the  technical literature turns out to be quite inefficient
when using real traffic traces over a  long period of time, we introduce a simple adaptive
scheme that closely  follows the variations of traffic.  When  tested against real traffic
traces, the proposed on-line algorithm performs well in the sense that the detection ratio
of  long flows  by the  algorithm  over a  long time  period  is quite  high.  Beyond  the
identification  of elephants,  this same  class of  algorithms is  applied to  the closely
related problem of detection of anomalies in  IP traffic, e.g., SYN flood due for instance
to attacks.  An algorithm for detecting SYN and volume flood anomalies in Internet traffic
is designed. Experiments show that an anomaly  is detected in less than one minute and the
targeted destinations are identified at the same time.
\end{abstract}

\section {Introduction}

\subsection*{Problem statement}
We address in this paper the  problem of designing an  on-line algorithm for  identifying  long flows in  IP
traffic.  From the point of view of
traffic engineering, this is an important issue.  This is
also an illustrative and
simple  example exhibiting  the importance  for on-line  algorithms to  adapt  to traffic
variations.  Traffic variations within a flow of packets may be due to several factors but the  way the TCP  
protocol  adapts the  throughput of  connections based  on the  congestion of  the network, notably through the number of packet losses, naturally leads to a stochastic behavior in the arrival patterns of packets. This is  a crucial issue which
is sometimes underestimated in the technical literature.  Moreover, as it will be seen in the second part, the methods developed for this problem
can be in fact used to design a quite efficient anomaly  detection algorithm.

An algorithm which can satisfactorily run   in some instances on limited traces can fail when handling a large traffic trace (e.g., several hours  of transit network IP
traffic) because of various reasons :
\begin{enumerate}
\item Performances deteriorate with time. The size of data structures increases without
bounds as well as the time taken by the algorithm to update them.
\item Poor performances occur even from the beginning.  Quite often, algorithms depend (sometimes in a
  hidden way  in the technical  literature) on constants  directly related to  the traffic
  intensity.  For a limited set of traces, they can be tuned ``by hand'' to get reasonable
  performances.  This  procedure  is,  however, not acceptable  in  the  context  of  an
  operational network. As a general requirement, it is highly desirable that the
  constants used  by algorithms automatically  adapt, as simply  as possible, to 
  varying traffic conditions.
\end{enumerate}

\subsection*{Identification  of  large flows}
It  is  well  known  that  if large   TCP  flows
(elephants) carry the main part of traffic  (in Bytes and packets), most of flows are small (in number
of packets).  A formal definition of ``long''  and ``small'' is defined later; as it will
be seen, this definition may depend on the  context. The discussion is kept informal until then.  Small
flows are typically generated by web browsing  while long flows are due to file transfers
(ftp, peer-to-peer  , etc.).  For  traffic engineering purposes like  billing, supervision
and security  for example,  it is important  to design  {\em on-line} algorithms  that can
identify some  of these long flows on the fly.

Due to  the very  high bit rate  and the  huge number of  flows in  IP traffic,  it is
unrealistic to maintain  data structures that can handle the set  of active flows. Indeed,
maintaining the  list of active  flows and  updating counters for  each of them  is hardly
possible in an on-line context. Consequently, only an estimation of the characteristics of
elephants can be expected within these constraints. 

A natural solution to cope with the huge amount of data in IP traffic is to
use hash tables.  A data structure using hash tables, a {\em Bloom filter},  proposed by
B.~Bloom~\cite{Bloom} in 1970,  has been used to test whether an element is a member of a
given set.  Bloom filters have been used in various domains: database queries,
peer-to-peer networks, packets routing, etc.  See Broder and Mitzenmacher~\cite{Broder} for
a survey. Bloom filters have been used by Estan and Varghese~\cite{Varghese} to detect
large  flows, see the discussion below.   
\nocite{Azzana:01} 

A Bloom  filter consists of  $k$ tables  of counters indexed  by $k$ hash  functions.  The
general principle is the following: for each table, the flow ID of a given packet (that is
the addresses and  port numbers of the source  and the destination) is hashed onto  some entry and
the corresponding counter is incremented by $1$. Ideally, as soon as a counter exceeds the
value $C$, it should  be concluded that the corresponding flow has  more than $C$ packets.

Unfortunately, since there is a huge number of small flows, it is very likely for instance
that a significant fraction (i.e. more than $C$ for example) of them will have the same
entry, incrementing the same counter, thereby creating a false large flow.
To  avoid  this  problem,  Estan  and Varghese~\cite{Varghese}  proposes  to  periodically erase
{\em all} counters. Without any a priori  knowledge on traffic (intensity,
flow arriving rate, etc.) which is usually the case in practical situations, the erasure  frequency can be either
\begin{enumerate}
\item too  low, and, in this case, the filters can be saturated:  Because of the large number of
  small flows, many of them may be   hashed on the same entry of the hash table and,
  therefore, the corresponding counter is increased accordingly,  and consequently
  creating a ``false'' large flow.   
\item too high and a significant fraction of elephants can be missed in this case: Indeed,
  the value of the counter of a given entry corresponding to a large  flow with a low
  throughput may not reach the value $C$ if the value of this entry is set to $0$ too often. 
\end{enumerate}
The  efficiency of  the  algorithm is therefore highly dependent on  the  period $T$  of
the  erasure  mechanism of the filters. This quantity is clearly related to the traffic
intensity. 

Starting from Estan and Varghese's algorithm,  an algorithm  based on  Bloom filters with
an additional  structure, the virtual filter, and a completely adaptive  refreshment
scheme is proposed.  As it will be seen, the proposed algorithm, based on simple
principles, significantly improves  the accuracy of algorithms  based on  Bloom
filters. Moreover,  the role  of  the constants  used by  the algorithm is thoroughly
discussed to avoid the shortcomings mentioned above.

\subsection*{Anomaly  detection}
An interesting application field of these methods is the detection of anomalous behavior, for instance due to  denial of
service. During such  an attack, a victim is  the target of a huge number  of small flows
coming from numerous sources connected to the network.  An on-line identification of such
anomalous behavior  is necessary for a network  administrator to be able  to react quickly
and to  limit the impact  of the attack  on the victim.  The main problem is in this case
to be able to separate  quantitatively ``normal'' variations of traffic from these sudden
bursts of traffic.   Here again,  adaptive properties of the detection algorithms  to
traffic conditions are essential to distinguish between normal variations of traffic and
attacks.  

Via an adequate  aggregation by   destination addresses,
the problem  is expressed in terms of  the detection  of  a single  large  flow. The
problem is then analogous to the one considered  in the detection of large flows: Most flows 
have to be quickly discarded so that only anomalous flows show up. 
Another algorithm using Bloom filters with an adaptive refreshment mechanism is also
proposed in this case: It is based on a fast refreshment scheme depending on the traffic
intensity  and on an adaptive estimation of some constants. This algorithm offers good
performances to detect SYN flooding attacks and also, via a variant, to detect more 
subtle (i.e. progressive) attacks such  volume flood attacks. 

The organization of this  paper is as follows: A detailed description
of the algorithm identifying large  flows is
given in Section~\ref{secalgo}. The algorithm proposed is tested against experimental data
collected from different types of IP networks in Section~\ref{secexpe}. The application to
the detection of denial of  service (DoS) attacks is developed in Section~\ref{secattack}.
Some performance  issues of the  algorithm are discussed in  Section~\ref{discussion}. Concluding remarks are presented in Section~\ref{conclusion}.

\section{Algorithms with Bloom Filters}\label{secalgo}

\subsection{Preliminary definitions}

In  this section,  we  describe the  on-line algorithm  used  to identify  large  flows  and
estimate their  volume.  Recall  that a flow  is the  set of those  packets with  the same
source and  destination IP addresses  together with the  same source and  destination port
numbers and  of the same protocol  type. In the  following, we shall consider  TCP traffic
only.

To simplify the notation, large  flows will be sometimes referred to as elephants and small flows as
mice. For several reasons, this dichotomy is largely used in the literature, see the
discussion in Papagiannaki \etal~\cite{Taft} for example. 
\begin{definition}[Mouse/Elephant] A mouse is a flow with less than $C$ packets. An elephant is a flow with at least $C$ packets.
\end{definition}
The constant $C$ is left as a degree of freedom in the analysis. Depending on the target
application, $C$ can be chosen to be equal to a few tens up to several hundreds of
packets. The choice of $C$ is left to the discretion of the operator. 

In this first  part, one investigates the  problem of on-line estimation of  the number of
elephants. This is probably the simplest  problem with all the main common difficulties in the design of  algorithms  handling Internet traffic: large order  of
 magnitudes and
reduced computing and memory capacities.

Note  that the estimation of  the {\em total} number of flows in  an
efficient and nearly optimal way is a quite different problem. Several
on-line algorithms have  recently    been
proposed by  Flajolet \etal~\cite{Flajolet}   and  Giroire   and  Fusy~\cite{Giroire}.
Unfortunately,  the corresponding algorithms are not able to identify
elephants as previously defined.

\subsection{Bloom filters}

\begin{figure}[htbp]
\scalebox{0.6}{\includegraphics{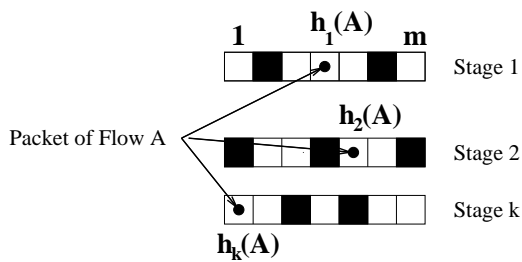}}
\caption{A Bloom filter \label{filter}}
\end{figure}

The starting point is the algorithm  based  on  Bloom filters designed
by Estan  and
Varghese~\cite{Varghese}.   The filter, see Figure~\ref{filter}, consists of $k$
stages. Each stage  $i\leq k$ contains $m$ counters taking values from $0$ to $C$. It is
assumed  that $k$ independent hash functions $h_1$, $h_2$, \ldots, $h_k$ are
available. The total size of the memory used for the filter is denoted by $M$, recall that
$M$ should be of the order of several Mega-Bytes.  An additional  auxiliary  memory is
used to store the identifiers  of detected elephants.  

The algorithm works as follows: All counters are initially set equal to $0$;  if a packet belonging to a flow $A$ is received then:\\
\noindent
--- If $A$ is in the memory storing elephant IDs then next packet.\\
\noindent
--- If not, let $\min(A)$ the minimum  value of the counters at the entries $h_1(A)$,
  \ldots, $h_k(A)$ of the $k$ hash tables. 
\begin{itemize}
\item  If $\min(A)<C$, all the corresponding counters having the value $\min(A)$ are incremented  by $1$.
\item  If  $\min(A)=C$, the flow of $A$ is added to the memory storing elephant IDs. The
  flow is detected as  an elephant.  
\end{itemize}
The algorithm as such is of course not complete since  small flows can be mapped repeatedly
to the same entries and create false elephants. One has therefore to clear the filters
from the influence of these undesirable flows. Estan  and Varghese~\cite{Varghese}
proposed to erase all counters of the filters on a periodic basis (5 seconds in their
paper):

\bigskip
\noindent
{\em Estan and Varghese's refreshing mechanism}
\begin{quote}
\begin{itemize}
\item  Every $T$ time units:\\
All counters are re-initialized to $0$. 
\end{itemize}
\end{quote}
Ideally, the constant $T$ should be directly related to the traffic
intensity.  On the  one hand, if
the refreshment mechanism occurs frequently, the counters corresponding to real elephants
are decreased too often and a significant fraction of them may not reach the
value $C$ and therefore many elephants will be missed. On the other hand, if $T$ is too large then,
because of their huge number, small flows may be mapped onto the same
entry and would increase the corresponding counter to the value $C$, creating a false
elephant. This periodic refreshing mechanism could perfectly work if there would be a way to change the
value of $T$ according to the order of magnitude of the number of small flows. Such a
scheme is however not easy to implement in practice. Moreover, in  Section~\ref{secexpe} experiments based on real traffic traces show that the periodic erasure scheme leads to a poor detection ratio of elephants.  This clearly exemplifies the fact that   the design of simple robust traffic adaptation scheme is not
an easy task in general (think again to the congestion avoidance mechanism of TCP). 

Our contribution to this algorithmic setting is two-fold: First, a refreshing mechanism of
hash tables  properly defined on the  current state of the  filters and not  bound to some
fixed time scale is proposed.  Second, an additional data structure,  the virtual filter, maintained
to get  a precise estimation  of the {\em  statistics} of these  large flows (and  not only
their number) is introduced.  These two aspects are  separately described in the following.

\subsection{Adaptive Refreshing Mechanism}
The general principle is the following: If the state of filters is declared as overloaded
then all positive counters are decremented by one.  

Note that  the values of the counters are only decreased by one
instead of reinitialized to
$0$. The idea is that, if the overload condition of  filters is properly chosen, then most of the
values of the non-zero counters will be low. Remember that from the structure of traffic, when compared
against  the number of mice, there are only a  few elephants. The key point  is that counters
corresponding to elephants will not be decremented  to $0$. This property is important if one wants to
accurately estimate the number of packets of elephants.
 
Two different criteria to declare when the state of the filter is overloaded are proposed. 
\begin{itemize}	
		\item RATIO criterion.  Define $r$ as the proportion of non null counters in the multistage
                  filter,  the filter is overloaded when $r$ is above some  threshold $R$
                  ($90\%$ for example). 
		\item AVERAGE criterion. Define $avg$  as the average of counters values. The filter  is overloaded
                  when  $avg$ is above some  threshold $AVG$ ($C/2$ for example). 
\end{itemize}
The adaptive property of the scheme proposed is clear: As long as the state of the filters
is not overloaded then nothing occurs and if there is a peak of activity,  the
filters are quite quickly filled and the refreshment mechanism is automatically executed. 

The rationale behind the RATIO criterion is that if  most of counters are non-zero then, very likely, mice
have contributed  to a significant fraction  of the values  of the counters so  that false
elephants  show up. Thus, this proportion must be bounded. The best
threshold is difficult to  find. Thus an interesting alternative is
the AVERAGE  condition, which considers the average value  of counters rather than  the number of
non-zero counters. Roughly speaking, this corresponds to the saturation threshold
of the filter. 
Notice the mean size of mice which is in practice around $4$, can
raise up to $7$ for some traffic types. In this case, even if the proportion
of non-null counters is significant, counters must be decremented more
often to
avoid accumulation of mice generating false elephants.  This is why
the AVERAGE  condition considers the average value  of counters. Our experiments show  nevertheless that condition RATIO  is sufficient
for most of IP traffic types.

Because the number of mice is much larger than the number of elephants, collisions between
elephants and mice can be neglected.  False elephants are mainly caused by collisions in
the  hash  table  between  short  flows.    Missing  elephants  is  the  drawback  of  the
algorithm. An elephant  having $f$ packets, $f \geq  C$, can be missed if  its counters do
not  reach the  threshold  $C$ because  of  the refreshment  mechanism  (all counters  are
decreased by one when the state of the filters is overloaded).

The number of entries in the memory storing elephants gives an estimation of their total
number. It is also possible to store  additional variables for each flow in this memory,
for instance the starting and finishing time of the elephant corresponding to the arrival
times of the first and last packets, the number of packets, the total volume
in bytes, the number of segments of a certain type (typically SYN segments for attack
detection), etc.  

\subsection{Virtual Filter}

Missed elephants can  be divided into two categories: elephants  with low throughput (less
than  the  refreshment  frequency) and  small  elephants.  An  elephant having a number of
packets slightly larger than $C$,  can then be missed  if  there  is at least  one
refreshment  during  its  life time.  The  following improvement of the algorithm aims at
reducing  the number of missed elephants by giving  elephants more chance to be
captured.  

The available  memory is  divided in two  halves.  In  the first half,  a Bloom  filter as
defined above   is implemented, it will be  called the virtual filter. It  operates exactly in
the same  way for  incrementing and refreshing  counters.  The second half  is another
Bloom filter, called the real filter;  its  counters are incremented in the same way as for
the virtual filter but no refreshment mechanism is used except that when a counter becomes
equal  to $0$  in the  virtual filter, in that case,  it is  also set
to $0$ in the real filter. 

The proportion of  non null counters is  thus the same for  the two filters. The  identification of
elephants is done  with the values of counters  of the real filter, when  all the counters
corresponding  to some  flow  are equal  to  $C$. Note  that since  the  counters are  not
decremented by one, it is less likely that  some packets of elephants will be lost in this
manner. The  value of a  counter in the  real filter is  therefore always higher  than (or
equal  to) the  corresponding counter  in  the virtual  filter. The  number of  identified
elephants is thus higher than what is obtained with the initial version of the algorithm. In
particular small elephants have more chance of being identified.

The drawback of the virtual filter is that, in some cases, it can introduce new false
positives. As the counters in the real filter are  higher, mice are more
likely to be considered as elephants. This especially happens when the
mean  size of mice
is not small enough compared to the threshold $C$.

\section{Experimental Results}\label{secexpe}
In this section, the efficiency of the  algorithm and the impact of some of its parameters
are   discussed.  

To evaluate the  performance of the algorithm, two different traces  have been tested: the
first  trace contains  commercial  traffic from  the  France Telecom  IP backbone  network
carrying ADSL traffic. This traffic trace has  been captured on a Gigabit Ethernet link in
October 2003 between 9:00 pm and 10:00 pm. This time period corresponds to the peak activity by
ADSL customers, its duration is $1$ hour  and contains more than 10 millions of TCP flows.
The             second            trace             ``20040601-193121-1'',            URL:
http://pma.nlanr.net/ Traces/ Traces/ long/ ipls/ 3/,  contains  academic  traffic issued  from
Abilene III.

\subsection{Results}

In our experiments, the  filter consists of 10 stages associated to 10 independent random
hash functions $(k=10)$. Elephants are here defined as flows with at least 20 packets
$(C=20)$. 

First we apply the algorithm proposed  by Estan and Varghese~\cite{Varghese} to the France
Telecom trace in order to identify elephants  for which the refreshment time period is set
to $5$ seconds as specified in that paper.  Recall  that this algorithm uses a periodic erasure scheme   of all  counters to  refresh the filter.   Results are  compared to  the adaptive
refreshment using  the RATIO criterion.   To be fair  in the comparison, at  a refreshment
instant, instead of decrementing them by one, all counters are set  to zero like in Estan
and Varghese algorithm.

\begin{figure}[htbp]
\begin{picture}(70,180)(10,20)
\put(-100,200){\scalebox{0.35}{\rotatebox{-90}{\includegraphics{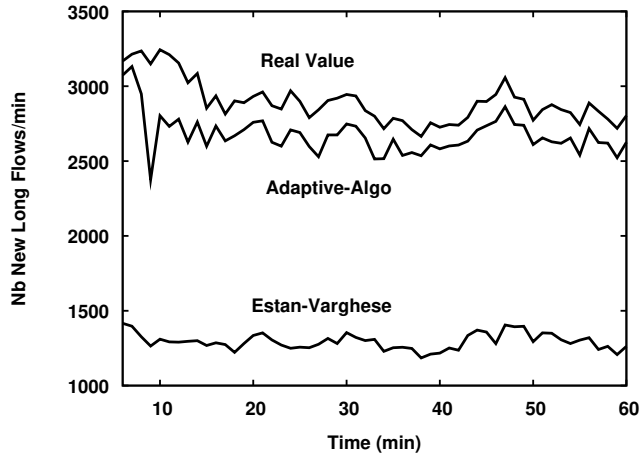}}}}
\end{picture}
\caption{Impact of the adaptive refreshment on the estimation of elephants number,
$M=1.31MB$, $R=90\%$, France Telecom trace \label{fig2}}
\end{figure}

The number  of new elephants per  minute found by the  algorithms and its  exact value are
plotted in Figure~\ref{fig2}. It shows that  the periodic refreshment of Estan and Varghese (5
seconds) is  not adapted  to the  traffic trace since  many elephants  are missed  in this
case. The refreshment frequency is too high and elephants cannot send their 20 packets in only
5 seconds.  This is due to the fact that in the ADSL traffic trace, elephants are
generated by peer to peer file transfers, which are basically with low bit rates (see Ben
Azzouna \etal~\cite{Nadia} for more details).

A change  of the value  of the period in Estan and Varghese's algorithm would probably
improve the  accuracy but it  is not clear how it  can be done ``on line''.   On a one
hour long traffic  trace, this parameter 
has to be in  fact changed regularly.  This is not necessary for  short traces, a few tens
of thousands of packets say, but this becomes an issue for long traffic traces. 

Using the
adaptive refreshment with a threshold $R=90\%$ and a small memory of size $M=1.31MB$, only
about $12\%$ of the elephants are missed.  With a memory size of $5.24$MB, the error is of
the order of $2$\%.  See Figure~\ref{fig5} below.

Another   important  feature  of   the  adaptive   algorithm  which   can  be   seen  from
Figure~\ref{fig2} is that it follows very closely the variations of elephant traffic, this
is also true for  Estan and Varghese algorithm but in a much  less accurate way.  This is,
in our view, the benefit of the adaptive property of our algorithm.

\begin{figure}[htbp]
\scalebox{0.32}{\rotatebox{-90}{\includegraphics{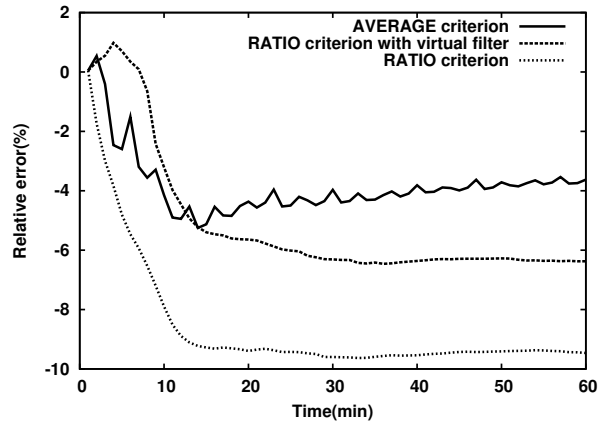}}}
\caption{Impact of refreshing mechanism and the virtual filter on the estimation of elephants number, $M=1.31MB$, $R=90\%$, France Telecom trace \label{fig3}}
\end{figure}

\begin{figure}[htbp]
\scalebox{0.32}{\rotatebox{-90}{\includegraphics{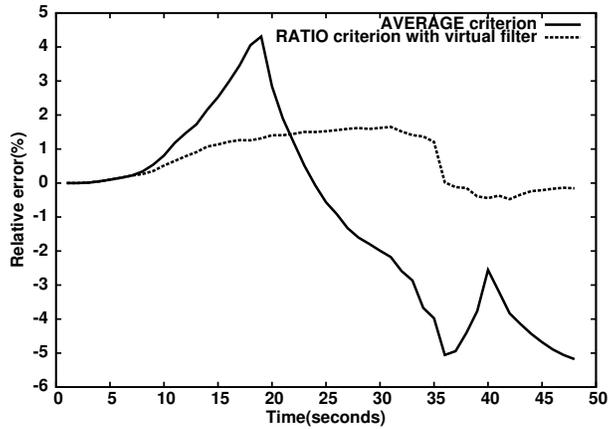}}}
\caption{Comparison of refreshing criteria  for Abilene trace with $M{=}1.31MB$ and $R=90\%$. \label{fig4}}
\end{figure}

Figure~\ref{fig3} gives the relative error on the estimation of the  number  of elephants for the 
three versions of the algorithm: with the refreshment using RATIO and AVERAGE criteria.
Both RATIO and AVERAGE criteria give accurate estimations of the total number of
elephants. The fact that the relative error remains under 7\% for all the duration of
the trace shows stability and robustness of the algorithm. The same experiments performed
on Abilene trace give similar results; see Figure~\ref{fig4}. So the adaptive algorithm is
an efficient method of refreshing  the filter without impacting too much the estimation of the
number of elephants. 

\begin{figure}[htbp]
\begin{picture}(70,180)(10,20)
\put(-100,200){\scalebox{0.35}{\rotatebox{-90}{\includegraphics{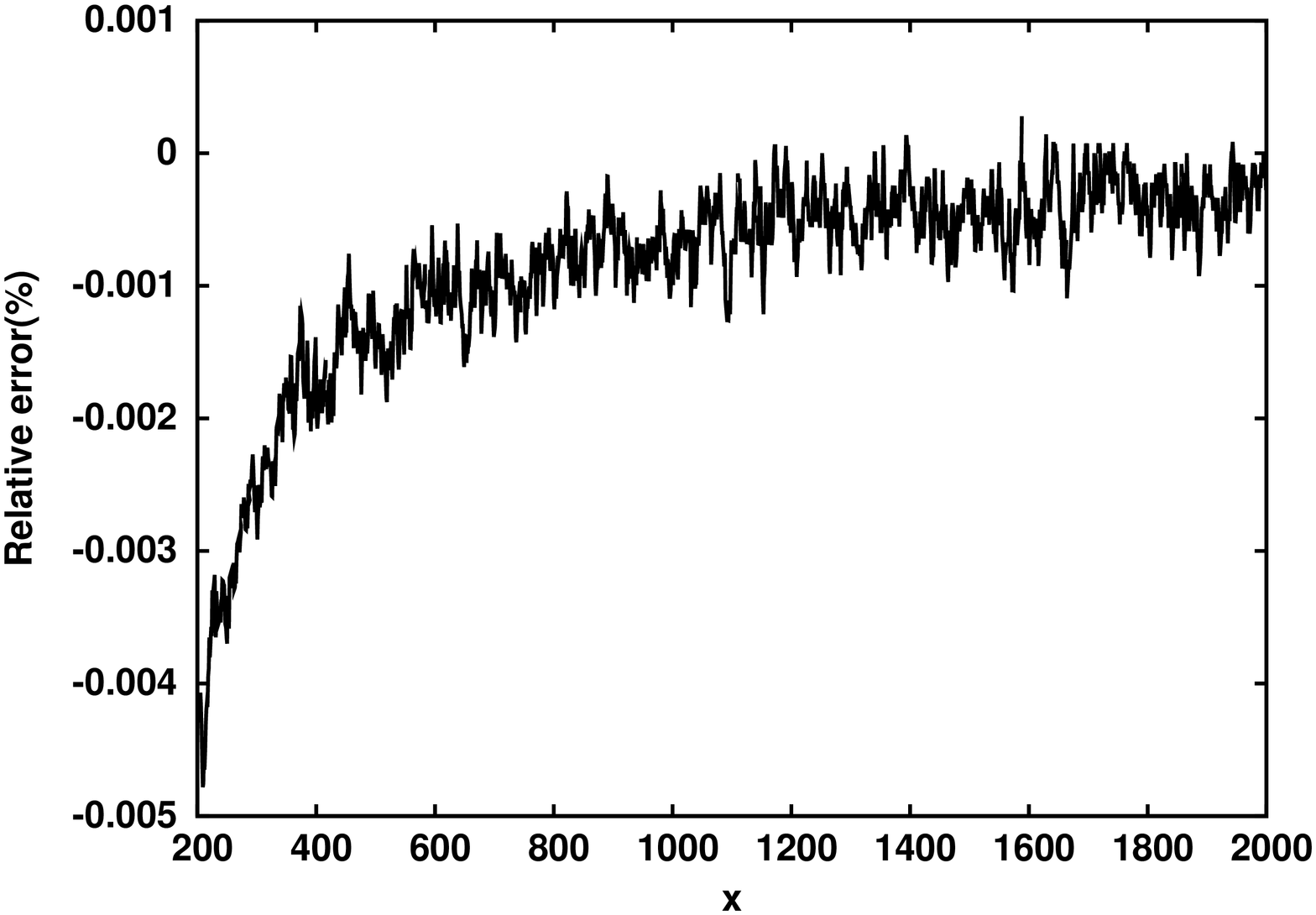}}}}
\end{picture}
\caption{Relative error on the number  of flows with more than $x$ packets, $M=1.31$MB, $R=90\%$. France Telecom trace\label{distrib1}}
\end{figure}

\begin{figure}[htbp]
\begin{picture}(70,180)(10,20)
\put(-100,200){\scalebox{0.35}{\rotatebox{-90}{\includegraphics{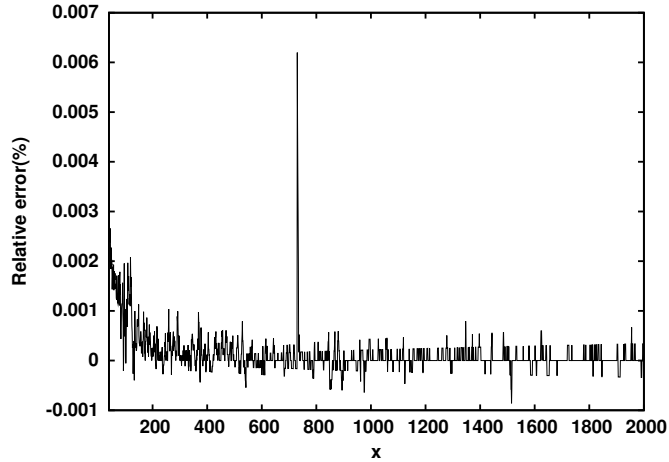}}}}
\end{picture}
\caption{Relative error on the number  of flows with more than $x$ packets, $M=1.31MB$, $R=90\%$. Abilene trace\label{distrib2}}
\end{figure}

Once an elephant has been  identified, it is registered in an auxiliary memory together with the
number of packets seen and each time a  packet of this flow is seen, this value is
incremented by 1. In this way, one can estimate the statistics of the sizes of elephants. 
Figures~\ref{distrib1} and \ref{distrib2} show that this statistics of this estimation of
the  number of packets per elephant is really very close to the real value for the two different
traces.  

\subsection{Impact of the  $M$ and $R$ parameters}

\begin{figure}[htbp]
\begin{picture}(70,180)(10,20)
\put(-100,200){\scalebox{0.35}{\rotatebox{-90}{\includegraphics{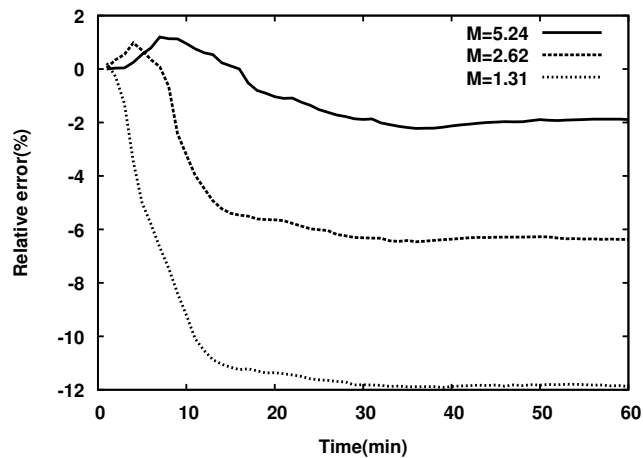}}}}
\end{picture}
\caption{Impact of memory size $M$ of the filter,  $R=90\%$ \label{fig5} }
\end{figure}

In Figure~\ref{fig5}, we analyze the impact of the size $M$ of the memory used for the Bloom filter on the estimation of the number of
the elephants. As expected, using a larger memory improves the accuracy. The error is
very close to zero with a memory size of  only  $5MB$. In fact 
the filter is refreshed less frequently which gives more chance for elephants to be
detected. 
    
\begin{figure}[htbp]
\begin{picture}(70,180)(10,20)
\put(-100,200){\scalebox{0.35}{\rotatebox{-90}{\includegraphics{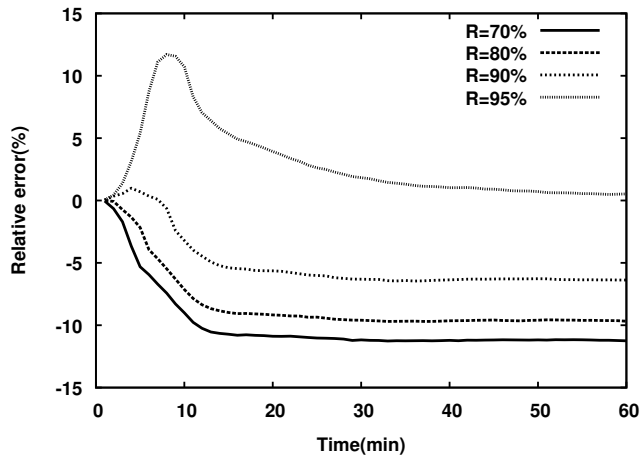}}}}
\end{picture}
\caption{Impact of  $R$ for RATIO criterion,  $M=1.31Mo$ \label{fig6}}
\end{figure}

Figure~\ref{fig6} shows the dependence of the accuracy of the estimate for several
values of  the threshold $R$. A threshold of $90\%$ gives a good estimation of the number of
elephants. We just miss about $7\%$ of the elephants. With a higher threshold, we miss
less elephants but some false positives can be added. So there is clearly a trade-off on the
choice of $R$. See  Chabchoub \etal~\cite{Chabchoub:02} for more details. 

\section{Anomaly Detection}\label{secattack}

\subsection{Context}
Several types of anomalies  are considered in this section in connection with   denial
of service (DoS) attacks.   Here  we are only interested  in
{\em SYN flood} and {\em volume flood} attacks which  are the most common DoS attacks. See
Hussain \etal~\cite{Hussain}  for   a  classification   of  DoS attacks. 

A {\em SYN flood} exploits a weakness in the connection phase of TCP, also called ``the three
way handshake''. This attack consists of sending a large number of SYN packets to the same
destination (or group of destinations) during a small interval of time. Due to the TCP
implementation, the destination  allocates resources to all these connection requests
and will maintain many half-open connections waiting for acknowledgments from sources for
about one minute. A large number of SYN packets  consume therefore a significant fraction
of the resources of the targets and, at the end, the corresponding machines become
unreachable  (see Wang  \etal~\cite{Wang} for more details). In this setting the goal is
to design  an one-line algorithm which can detect an attack in less than one minute. Such a
detection can be used by the network operator in order to filter SYN segments towards the
victim.  

While  a  SYN flood consists of a sudden arrival  of a large number of SYN segments, 
a {\em volume flood} attack uses a few TCP flows and gradually transmits with a steady increase
of the transmission rate  a huge amount of data which will consume the available bandwidth
of the target.  

Several methods  have been developed in  the technical literature for  DoS flooding detection; they
are mainly based on TCP properties such as periodicity in Chang \etal~\cite{Cheng}, or SYN
and  FIN  packets  counting   in  Wang  \etal~\cite{Wang},  Barford  \etal~\cite{Barford},
Krishnamurty  \etal~\cite{Krishna}.  Most of  them suffer  from scalability  or robustness,
especially if  only sampled traffic  is available  as it is  usually the case  in backbone
networks.

For SYN  flood detection, the  main difficulty is  in distinguishing between  the (normal)
variations of traffic  and a sudden and anomalous sequence of SYN  packets. In the technical literature,
attacks  are sometimes  defined  as a  notable  variation from  the  standard behavior  of
specific parameters  of the model:  parameters of some specific  statistical models or
long range dependence variables like Hurst parameters for  signal processing approaches
for example. If the algorithms based on these representations may be efficient to detect some anomalous
behaviors, they  cannot, in general, assert the  nature of the attack  because they handle
a aggregated information on the flows rather than a more detailed description of the
traffic.  See Chatelain \etal~\cite{Borgnat} and Lakhina \etal~\cite{Diot}.

\subsection{SYN flood attacks}

The  algorithm proposed for an on-line detection of SYN and volume flood is derived  from
 the algorithm presented  in Section~\ref{secalgo}, but with a different refreshing 
mechanism of the Bloom filter. 

As explained above, SYN packets  with a given  destination  address are  aggregated  as a  single 
``flow''. In this case, by using a Bloom filter as before,  the refreshing mechanism of the
multistage filter has a different purpose: it should eliminate quickly all normal flows
using an aggressive refreshing mechanism so that if a ``large''  flow survives then it must
be a SYN flood attack. As it is easily seen, the term ``large''  has to be properly
defined. Roughly speaking, this means that such a flow  is much larger  than the other ``normal'' flows. 
Again, because of the variation of traffic, an adaptive scheme has to be devised to properly define
 these concepts. 

The main idea of the algorithm is to  evaluate a varying average $m_n$ of the largest flow
in several sliding time windows of  length $\Delta$. The quantity $m_n$ describes ``normal
flows''; it is periodically updated in order to adapt to varying traffic conditions. It is
a weighted average that takes into account all its past values to follow carefully traffic
variations but  not too closely. If  a flow in the  $n$th time window is  much larger than
$m_{n-1}$, it is considered  as an attack, and the moving average  is not updated for this
time window.

The following variables are used. 
\begin{itemize}
\item As before, $r$ is the proportion of non-zero counters in the Bloom filter. 
\item $S$ is a multiplicative detection threshold. Roughly speaking, an attack is declared
when an observation is $S$ times greater than the ``normal'' behavior. The value of $S$ is fixed  by
the administrator. 
\item $R_s$ and  $R$ are thresholds for the variable $r$. The constant $R_s$  is 
 independent of traffic and  taken once and for all  equal to $50$\%  and $R$ is a
 variable threshold depending on the traffic type considered.  
\item $\alpha$ is the updating coefficient for averages,; $\alpha = 0.85$ in our experiments. 
\item $\Delta$ is the  duration of the initialization phase ($1$ mn in the paper). It is in fact a bound for the time before which
an attack should be detected.
\item $m_n$ is the weighted moving average for the $n$th time window.
\end{itemize}
The algorithm starts with an initialization phase of length $\Delta$ in order to evaluate
the threshold $R$. At the end of this phase,
$R$ will be definitively fixed for the rest of the experiment.  In addition, as this phase
corresponds to the first time window, the moving average $m_1$ will be initialized as the
biggest counter obtained.  See Table~\ref{alogattack} for the description of the algorithm.

\begin{table}
\hrule
Initialization phase:
\begin{itemize}
\item All counters are $0$. 
\item The Bloom  filter is progressively updated with SYN packets by using their
  destination address. 
	\begin{itemize}
	\item After a duration $\Delta$, evaluate the variable $r$
		\begin{itemize}
		\item if $r \leq R_s$ then $R:=r$ else $R:=R_s$.
		\item  $m_1 :=$ maximum of the values of counters of the multistage filter.
		\end{itemize}
	\end{itemize}	
\end{itemize}	
Detection phase: the $n\ th$ time window
\begin{itemize}
\item At the beginning all counters are initialized to $0$. 
\item The Bloom  filter is progressively updated with SYN packets by using their   destination address. 
	\begin{itemize}
	\item if a counter exceeds $S\ m_{n-1}$, an attack is declared.
	\item if $r \geq R$
		\begin{itemize}
		\item $\max_n :$ maximum of the values of counters of the multistage filter.
		\item if $\max_n<S\ m_{n-1}$ \\
        $m_n = \alpha\ m_{n-1}+(1-\alpha)\max_n$
        \item start the $(n+1)$th time window.
		\end{itemize}
	\end{itemize}	
\end{itemize}
\hrule
\caption{Algorithm for SYN flood detection.\label{alogattack}}
\end{table}

Note that an alarm is declared during the $n$th time window when the value of a counter is
greater than $Sm_n$. At the beginning, the first time
window is fixed (its duration is $\Delta$) but, since the evolution depends on the occupation rate of
the filters, the duration of the other time windows is variable.  If traffic characteristics are not much varying,
time windows durations remain around one minute. In this case, an attack is detected
at the latest after one minute  so that the network administrator can react quickly.

\subsection{Volume flood attacks}

For  progressive attacks, the impact on  traffic cannot be clearly seen in
a time window of one minute. In fact the attack can be so slow that it could be  locally considered
as a normal traffic variation. This kind of attacks has typically a long duration. In this
situation, we  consider a larger time window in order to detect the anomalous impact of the
attack on  traffic. Thus,  to cope with these attacks, the algorithm
is used but with  a larger time window $\Delta'$ of 5 minutes. This new filter operates
in the same way but on a longer time scale and is completely independent of the first
filter. In particular, it  has its own parameters: $R'$, $r'$, $R'_s$, $m'_n$, $\max'_n$, and  $S'$. 

\subsection{Experimental Results}
To evaluate and validate the attack detection algorithm described in the previous section,  we run experiments with  two
France Telecom traces, one from the IP collect network carrying in majority ADSL traffic and the other from the IP transit network (OTIP). In this latter case, only
sampled traffic is available. The characteristics of the traffic traces are given in Table~\ref{charac}.   

\begin{table}[hbtp]
			\begin{center}
			\begin{tabular}{lccc}
			\hline
			Traces	 &  Nb. IP pack. &  Nb. Flows& Duration\\
			\hline
			OTIP	&$105.10^6$	&$4.10^6$	&~3 days\\
			ADSL &$825.10^5$	&$32.10^5$ 	&3 hours\\
				
			\hline
			\end{tabular}
			\end{center}
			\caption{Characteristics of sampled traffic  traces used for attack detection. \label{charac}}
\end{table}


To detect SYN and volume flood using two time scales ($\Delta=1$ mn and $\Delta'=5$ mn), we need
four filters. Each filter contains ten stages $(k=10)$ and has a total size $M$ around
$1MB$. 

\begin{figure}[hbtp]
\scalebox{0.32}{\rotatebox{-90}{\includegraphics{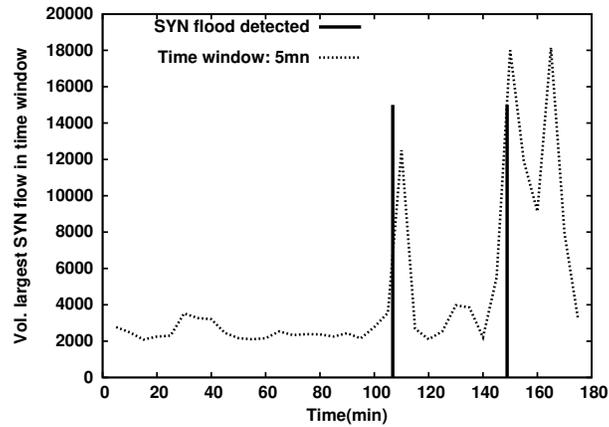}}}
\caption{SYN flood detection for ADSL  trace with $S{=}5$ and $S'{=}3 \label{fig7}$}
\end{figure}

In Figure \ref{fig7},  the ADSL trace is  divided into several time windows  of $5$ mn and,
for each interval, the volume of the largest SYN flow is computed. The observed peaks seem
to correspond to  attacks. Tested on this trace, the algorithm  detected two SYN flood
against two different IP addresses. The  response time of the algorithm is satisfactory as
the alarms are  raised at the beginning of  the attacks. It should be noted  that when the
duration of the time window is $1$ mn, only the second attack is detected.

\begin{figure}[hbtp]
\scalebox{0.32}{\rotatebox{-90}{\includegraphics{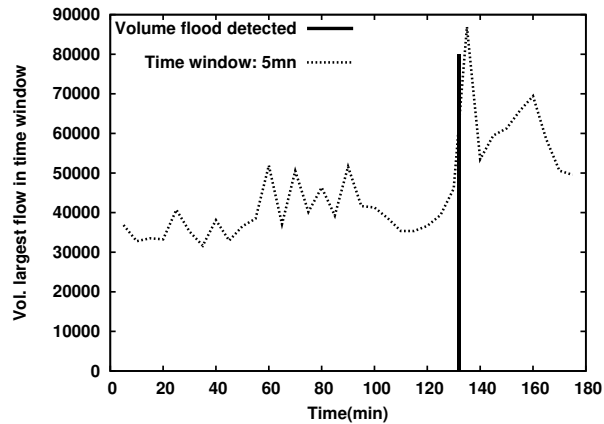}}}
\caption{Volume flood detection for ADSL trace with $S{=}5$ and $S{'}{=}2$ \label{fig8}}
\end{figure}

In Figure \ref{fig8}, the same trace is used to detect volume flood. The
volume of the flow is now the number of packets which are not SYN packets. SYN packets are not computed
to prevent from considering some SYN flood as volume flood. The algorithm detects one
volume flood using the time window of $5$ mn. When the duration of the time window
$1$ mn, no attack is detected. 

\begin{figure}[hbtp]
\scalebox{0.32}{\rotatebox{-90}{\includegraphics{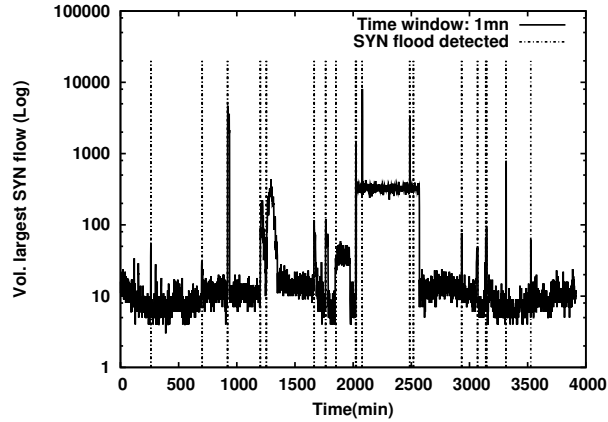}}}
\caption{SYN flood detection for OTIP trace with $S=5$ and $S'=3$ \label{fig9} }
\end{figure}

\begin{figure}[hbtp]
\scalebox{0.32}{\rotatebox{-90}{\includegraphics{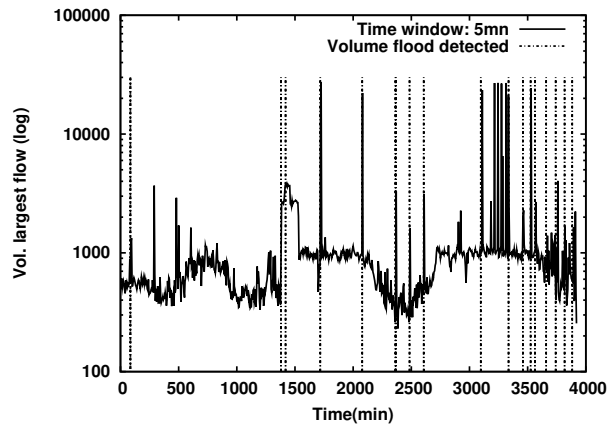}}}
\caption{Volume flood detection for OTIP trace with $S=5$ and $S'=2$ \label{fig10}}
\end{figure}

In Figures~\ref{fig9} and \ref{fig10} the OTIP trace is considered. This trace contains
many attacks. As it can be seen, the algorithm raises several alarms which coincide with
the largest flows represented by the highest peaks. 
 
\subsection{Remark on thresholds}

The algorithms use  the variables $S$ for SYN  flood and $S'$ for volume  flood.  They are
related  to  the network  administrator's  decision about  the  precise  definition of  an
anomalous behavior.   In the experiments  with the OTIP  trace for SYN flood  detection or
volume  flood detection,  there is  clearly a  set of  events which  will be  qualified as
``attacks'' for a large range of values of $S$ and $S'$. Note however that, for some large
but ``milder'' variations, the qualification as attack will depend on the particular value
of these parameters. There is no way to avoid this situation in our view. This is the role
of the administrator to define the level of abnormality in traffic.

\section{Performance issues} \label{discussion}
In  this section,  we  discuss  briefly, from  a  modeling point  of  view,  some of  the
performance issues concerning the refreshing  mechanism of the algorithms presented in the
paper.      They     are      addressed      in     more      detail     in      Chabchoub
\etal~\cite{Chabchoub:02,Chabchoub-3}.   Recall  that our  algorithms  have the  following
parameters:
\begin{itemize}
\item $R$: overload ratio of the filter,
\item $C$: minimum size of large flows (elephants), 
\item $m$: number of counters per hash table,
\item $k$: number of hash tables in the filter.
\end{itemize}
The number  $m$ is large  and the value  of $C$ is  fixed.  The main  issue is in  fact to
investigate the sensitivity  of the value of  $R$ on the performances: How  can one choose
$R$ no too large to avoid as much  as possible false elephants but large enough to prevent
from missing many true elephants.  The  problem reduces to estimate the error generated by
mice, i.e., the ratio of false positives  in a simplified model where there are just mice.
The case where $k=1$ is first  investigated as the simplest model.  Then simplified models
are developed for the case where $k\geq 2$.

 In a first step, a one-stage filter  is considered and traffic is supposed to be composed
 of only  of mice of size~$1$.  The  analysis uses Markovian techniques:  If $W_n^m(i)$ is
 defined as  the proportion of counters  with values $i$, $i  \in{0,\ldots,C}$ just before
 the $n$th refreshment for a filter with $m$ counters then the process
\[
\left(W_n^m,     n\geq
0\right)=\left(\left(W_n^m(i), i=0,\ldots,C\right),  n\geq 0\right)
\]
 is a Markov  chain on a finite  state space with invariant distribution  $\pi_m$.  As $m$
 gets large,  it is shown that the  sequence $\left(W_n^m, n\geq 1\right)$  converges to a
 dynamical system with  unique fixed point $\overline w$. It turns  out that $\overline w$
 has  a nice  interpretation  in terms  of  the stationary  measure  $\mu_{\lambda}$ of  a
 $M/G/1/C$ queue with service time $1$ and arrival rate $\lambda$ and
\[
\overline  w=\mu_{\lambda(\overline  w)}
\]
where $\lambda(w)=\log(1+w(1)/(1-r))$.   As the  invariant measure $\mu_{\lambda}$  can be
computed as the solution of a linear system of $C+1$ equations, $\lambda(\overline{w})$ is
thus the solution to a fixed point equation.

The behavior of the system can be described as follows. At equilibrium, the average number
of  packets between  two refreshing  instants is  of the  order  of $\lambda(\overline{w})
m$.  Due to  finite capacity  $C$, this  quantity is  greater than  the number  of removed
packets  $Rm$   at  each  refreshment.   In  particular   $\lambda(\overline{w})$  is  not
necessarily   less  than   $1$.  For   $R$  enough   close  to   $1$,  it   is   shown  that
$\lambda(\overline{w})$ can in  fact exceed $1$ which changes  the qualitative behavior of
the  system:  if  the  arrival   rate  $\lambda(\overline{w})$  is  less  than  $1$,  then
$\overline{w}$  is concentrated  on  small  values $0,\;1,\;2\ldots$  of  the state  space
$\{0,1,\ldots  C\}$.  When  $\lambda(\overline{w})>1$ the  distribution  $\overline{w}$ is
mainly concentrated on the highest values $C$ and $C-1$. Since false positives are closely
related  to the  quantity $\overline{w}(C)$,  this implies  that the  proportion  of false
positives is much higher in this case.  Consequently, there is a critical value $r_c<1$ of
$R$,  which corresponds  to  $\lambda(\overline{w})=1$,  so that  the  performances of the
algorithm  deteriorate when $R>r_c$. Similar conclusions hold also in the case of
$k$-stage Bloom filter.

In practice, in the (simplified) case of $1$-packet mice, the critical rate $r_c$ is close
to $1$. But, for general mice size distribution, $r_c$ is much lower than $1$. Thus, the
RATIO criterion does not perform well for any value of $R$. To conclude, the analysis
confirms that the threshold $R$ has to be chosen, otherwise the RATIO criterion cannot
control the saturation of the filter. This control is contained in the  design of the AVR
criterion. 

Let us give an insight into  a model taking into account the case where just the counters
having the minimum value are incremented. The analysis can be generalized to  more
general situations. The idea is also to express quantities via a continuous time
process. The main tool is the system of $m$ queues with a Poisson arrival process with
rate $\lambda m$ where customers join the shortest of the $k$ queues chosen at random
among $m$. This system is well-known in the literature (see Mitzenmacher
\cite{Mitzenmacher-1}, Vvedenskaya \cite{Vvedenskaya-1}, Graham \cite{Graham-3} and
others). In order to use it, the model must be  slightly modified: the mouse increments
just one counter having the minimum value and $C$ is taken as $20/k$. The conclusion is
that the  behavior of the model should follow the same lines but many points are more
difficult  to catch. For example, as far as we know, the system of queues corresponding to
mice with general size distribution has never been studied in the
literature. Nevertheless,  the analysis gives rise to related models which could lead to
improve or simplify the algorithm.

\section{Concluding remarks}
\label{conclusion}

We have presented  in this paper an original adaptive  algorithm for identifying elephants
in Internet traffic. As earlier proposed by Estan and Varghese, this algorithm is based on
Bloom  filters, but instead  of periodically  erasing the  filter, we  introduce different
original criteria to decrement the various counters of the filter. In order to improve the accuracy
of the  algorithm, we have  introduced the concept  of virtual filter, whose  counters are
less  frequently decreased.  The  proposed  algorithm has  been  tested against  different
traffic traces and performs better than the one by Estan and Varghese.

Finally, the  proposed elephant  identification mechanism  has been adapted  in order  to detect
flood  anomalies (SYN  and volume  floods) in  Internet traffic.  This gives  rise to  a new
algorithm,  whose  key  parameters adapt  to  network  traffic.  This algorithm  has  been
successfully  tested  with   two  types  of  traffic traces  (corresponding  to
residential  and transit traffic).

\providecommand{\bysame}{\leavevmode\hbox to3em{\hrulefill}\thinspace}
\providecommand{\MR}{\relax\ifhmode\unskip\space\fi MR }
\providecommand{\MRhref}[2]{%
  \href{http://www.ams.org/mathscinet-getitem?mr=#1}{#2}
}
\providecommand{\href}[2]{#2}

\end{document}